\documentclass[iop, apj, tighten]{emulateapj}
\usepackage{apjfonts, soul}

\slugcomment{Final version 2016-04-18}

\shorttitle{Power Beaming Leakage Radiation As A SETI Observable}
\shortauthors{Benford \& Benford}

\begin{document}

\title{Power Beaming Leakage Radiation As A SETI Observable}

\author{James N. Benford}
\affil{Microwave Sciences, 1041 Los Arabis Lane, Lafayette, CA 94549 USA}
\and
\author{Dominic J. Benford}
\affil{NASA's Goddard Space Flight Center, Observational Cosmology Laboratory, Greenbelt, MD 20771}
\email{jimbenford@gmail.com}

\begin{abstract}
The most observable leakage radiation from an advanced civilization may well be 
from the use of power beaming to transfer energy and accelerate spacecraft. Applications suggested for power beaming involve launching spacecraft to orbit, raising satellites to a higher orbit, and interplanetary concepts 
involving space-to-space transfers of cargo or passengers.  We also quantify beam-driven 
launch to the outer solar system, interstellar precursors and ultimately starships. 
We estimate the principal observable parameters of power beaming leakage. Extraterrestrial civilizations would know their power beams could be observed, and so could put a message on the power beam and broadcast it for our receipt at little additional energy or cost. By observing leakage from power beams we may find a message embedded on the beam.
Recent observations of the anomalous star KIC 8462852 by the Allen Telescope Array (ATA) 
set some limits on extraterrestrial power beaming in that system. We show that 
most power beaming applications commensurate with those suggested for our solar 
system would be detectable if using the frequency range monitored by the ATA, and 
so the lack of detection is a meaningful, if modest, constraint on extraterrestrial power beaming 
in that system.  Until more extensive observations are made, the limited observation 
time and frequency coverage are not sufficiently broad in frequency and duration 
to produce firm conclusions. Such beams would be visible over large interstellar distances. This implies a new approach to the SETI search: Instead of focusing on narrowband beacon transmissions generated by another civilization, look for more powerful beams with much wider bandwidth. This requires a new approach for their discovery by telescopes on Earth. Further studies of power beaming applications should be done, which could broaden the parameter space of observable features we have discussed here. \end{abstract}

\keywords{space vehicles, extraterrestrial intelligence, stars:individual --- KIC 8462852}


\section{OBSERVABLE POWER BEAMING}

The most observable leakage from an advanced civilization may well be from the 
use of power beaming to transfer energy and accelerate spacecraft, both within 
and beyond the star system where the civilization is located.  In future, such 
applications may make the Earth's radiation in the microwave, millimeter and visible/near-IR 
parts of the electromagnetic spectrum be very intense. Beaming of power for a variety 
of space applications has been a frequent topic of study because it has many advantages. 
Beaming power for space transportation can involve Earth-to-space, space-to-Earth, 
and space-to-space transfers using high-power microwave beams, millimeter-wave 
beams or visible/near-IR lasers. Applications include launching spacecraft to orbit 
or raising satellites to a higher orbit.  Several investigators have studied interplanetary 
cargo transfers by beam-driven sail craft using radiation pressure, principally 
space-to-space commerce, launch into the outer solar system, and interstellar precursor probes starships. 
Reviews of power beaming applications \citep[Ch. 3]{Benford2008, Benford2013, Benford2016} 
 provide details on these applications, which would be superfluous to repeat
here.  Other means of reaching high speeds are rockets: fusion rockets, anti-matter rockets, Bussard ramjets. Alternative methods of propulsion are compared to power beaming in \citet{Moeckel72, Cassenti82, Dyson82, Matloff05}.  There is increasing agreement that power beaming is the most likely way forward.

The power levels are high, focused, and transient and could easily dwarf any of 
our previous leakage to space. These are not SETI \textit{signals} so much as \textit{leakage}, 
a detectable aspect of advanced civilizations. Studies have shown that leakage 
of TV and radio broadcast signals are essentially undetectable from one star to 
another, due to faintness and incoherence \citep{Sullivan1978}.  Planetary radars are 
stronger, but very transient in time and solid angle \citep{Billingham2014}. 
However, the driving of spacecraft by intense beams of radiation is far more focused 
than communication signals, more likely to repeat, and of course far more powerful.  Therefore they could 
be far more easily detected. 

It has previously been noted that such leakage from other civilizations could be 
observable \citep{Benford2008}. \citet{Guillochon2015} have quantified leakage from 
beaming for interplanetary space propulsion, its observables, and implications 
for SETI. Extraterrestrial Intelligence (ETI), having done the same thinking, could 
realize that they could be observed. Hence there may be a message on the power 
beam, delivered by modulating it in frequency, amplitude, polarization, phase, 
etc., and broadcast it for our receipt at little additional energy or cost. By 
observing leakage from power beams we may well find a message embedded on the beam.

We quantify the various classes of power beaming applications/missions, estimate 
the principal observable parameters and discuss the implications of observability 
of ETI power beaming leakage and our own future emissions.

\section{POWER BEAMING MISSIONS AND SPECIFICATIONS}

\subsection{Launch to Orbit} 

Beamed power can be used to launch spacecraft into orbit. \textit{Microwave 
thermal thrusters} operate on an analogous principle to nuclear thermal thrusters 
and have been experimentally demonstrated \citep{Parkin2003}.  In this concept, 
 high power microwave 
(HPM) beam radiates power to a thermal propulsion system in a single stage rocket. 
The spacecraft has a flat aeroshell underside covered by a thin microwave absorbing 
heat exchanger made of, for example, silicon carbide. The exchanger consists of 
$\sim 1000$ small channels carrying fuel, such as hydrogen, to the motor. 
To be specific, a system with, for example, a beam power of 300\,MW radiates from an aperture spanning 300 m impinging 
on a $7 \,{\rm m}^2$ converter on the underside, heating the hydrogen fuel, 
which exits the nozzle, Such rockets are much more efficient than conventional 
fuel-burning rockets.  By using HPM, the energy source and all the complexity that 
entails remains on the ground, and a beamed power transmission system carries 
the energy to the craft.  It has a high acceleration ascent trajectory, which provides 
most of the transfer of energy at short range, in order to minimize the size of 
the radiating aperture. For such a system the launch cost could potentially fall 
to as low as a few times the \textit{energy} cost of launch (as opposed to capital 
cost of a throwaway rocket and fuel), so low-cost and reusable launchers are possible. 

Frequencies used for power beaming depend on the location of the transmitter. For 
launch to orbit in an Earth-like atmosphere the microwave frequency window at about 
$1-10$\,GHz would be appropriate because atmospheric losses are low. For an Earth-like 
atmosphere with some oxygen and low water vapor, high-altitude sites, further windows 
are at 35\,GHz, $70-115$\,GHz, $130-170$\,GHz and $200-320$\,GHz.  With a different planetary 
atmosphere and weather patterns, different frequencies could make better economic 
sense and the absorption and breakdown thresholds would be different.

Launch rate on Earth could be as quick as one every three minutes per facility, based 
on the time required to accelerate a payload to low Earth orbit.  Launches could 
be bunched together to propel many craft in the space of a few hours to save on 
range operating costs. Additionally, a night launch is preferred for better beam 
propagation due to lower wind speeds and fewer clouds. Thus there could be a correlation 
of microwave intensity with the day-night cycle of the planet with pulse lengths 
that are $1-10$ minutes in duration.  There could be a correlation between the carrier 
frequency and planetary atmosphere type, as well.  Perhaps there may also be clues 
as to the type of microwave generating technology used in the linewidth and its 
frequency stability, once Doppler shift is accounted for.

\subsection{Orbit raising}

A lower power application of power beaming is orbit raising, where microwave energy from the ground is used to lift a satellite 
gradually into a higher orbit, is a lower power application of power beaming. An 
orbital transfer vehicle shuttles cargo from low Earth orbit (in the example above, 
taking the cargo from the thermal rocket, which could then be returned to the surface) 
to geosynchronous or cislunar orbits. The \citet{Brown1992} concept for this has 60-ton mass, with 
payload mass fraction of about 40\%.  The microwave beam of 10 MW 
provides electric power directly via a rectifying antenna to drive ion thrusters 
on the platform at a steady acceleration of $10^{-2}\,{\rm m/sec}^2$. 
 Such orbit-raising takes about half a year.  It is a good example of the strengths 
of power beaming: it is efficient, reusable and inexpensive, and can operate around 
the clock, although any given target can receive power and accelerate only when 
it is above the horizon of the transmitter. Efficiency improves by using a higher 
frequency, such as 94\,GHz, a water window for low-loss transmission. 

A ground-based or orbiting transmitter can impart energy to a \label{OLEHLINK18}satellite 
if they have resonant paths \citep{Benford2006} -- that is, the power beam source and satellite come near each other, either by waiting for the satellite to be overhead of the ground transmitter, or for both to be nearby while in orbit in space). When resonance occurs, and amount of energy specific to that particular conjunction is radiated to the satellite. Resonant 
orbit boosting, accelerating lower down in the gravity well and therefore nearer 
to the beam director is also more efficient due to the Oberth effect (a powered flyby in which a spacecraft falls into a gravitational well, and then accelerates when its fall reaches maximum speed, producing a greater gain in kinetic energy as compared to using the same impulse outside of a gravitational well). Such resonant orbits can use several transmitter locations.  The total time to escape Earth's gravity well can be as little as 10 days. Such transmitters will require powers up to $100\,$MW, but will be used a fraction of each satellites' orbital period, which will gradually lengthen.

\subsection{Interplanetary logistics}

A number of higher-velocity power beaming applications have been quantified for 
fast transit of the solar system -- Mars, Jupiter, Kuiper Belt, Plutinos, 
Pluto and the Heliopause. An attractive interplanetary mission could be the rapid 
delivery of critical payloads within the solar system.  For example, such emergencies 
as crucial equipment failures and disease outbreaks can make prompt delivery 
of small mass payloads to, e.g., Mars colonies, an imperative. Lasers or microwaves 
accelerate such urgent cargo with sail spacecraft at fast boost for a few hours 
of propulsion to speeds of $100-200$\,km/sec. The craft then coasts at constant high 
speed until decelerating for a few hours into Mars orbit (probably by a decelerating 
beam system like the one which launched it), giving a 10-day transit time \citep{Meyer1985}. This method has been extended to missions with 5\,gee acceleration near Earth 
\citep{Benford22006}. Using a ground station, acceleration occurs 
for a couple of hours for a 100\,kg payload. \citet{Guillochon2015} have quantified 
a strategy for detecting leakage transients from such ETI interstellar logistics. They estimate that if we monitor continuously, the probability 
of detection would be on the order of 1\% per planetary conjunction event. They 
state that ``for a five-year survey with $\sim 10$ conjunctions per system, 
about 10 multiply-transiting, inhabited systems would need to be tracked to guarantee 
a detection'' with our existing radio telescopes.

\begin{deluxetable*}{cccccc}[hbt]
\tablecolumns{6}
%\tabletypesize{\scriptsize}
\tablecaption{Representative Parameters for Applications of Power Beaming\label{table1}}
\tablewidth{0pc}
\tablehead{
\colhead{Application} & \colhead{Frequency} & \colhead{Power} & \colhead{Duration} & \colhead{Repeat Time} & \colhead{Beamwidth}\\
\colhead{} & \colhead{f} & \colhead{$P$} & \colhead{$T$} & \colhead{} & \colhead{$\Delta$$\theta$=2.44$\lambda$/D} \\
\colhead{} & \colhead{[GHz]} & \colhead{} & \colhead{} & \colhead{[sec]} & \colhead{[radians (arcsec)]} }
\startdata
Launch to Orbit & 94 & 300 MW & minutes & Immediate & $2\times 10^{-5}$ ($4.1''$)\\
Orbit Raising & 94 & 300 MW & hour & hour & $2\times 10^{-4}$ ($41''$)\\
Interplanetary & 68 & 0.3 TW & hours & Immediate & $4\times 10^{-6}$ ($0.8''$)\\
0.1 c Starship & 100 & 1 TW & 10 hours & days & $2\times 10^{-8}$ ($0.004''$)\\
0.5 c Starship & $3\times10^{14}$ \,Hz (1 $\mu$m) & 100 TW & years & years & $2\times10^{-11}$ ($4.1\mu$as) 
\enddata
\end{deluxetable*}

\subsection{Interstellar Probes}

Interstellar Probes are solar/interstellar boundary missions out to $\sim 1000$\,AU.
The penultimate is the interstellar precursor mission.  For this mission class, 
operating at high acceleration, the sail size can be reduced to less than 100\,m 
and accelerating power $\sim 100$\,MW focused on the sail. At 1\,GW, sail 
size extends to 1\,km and super-light probes reach velocities of 300\,km/s (63\,AU/year) 
for very fast missions of about 10 year duration \citep{Benford22006}. 

\subsection{Starships: the biggest and grandest missions}

Concepts of this sort require very large transmitter antenna/lens
and sail (e.g., $1,000\,$km diameters for missions to 40\,ly).  A Space Solar Power (SSP; see below) station radiates a microwave beam to 
a perforated sail made of carbon nanotubes with lattice scale less than the microwave 
wavelength.  The sizes of the first mission concepts were enormous, with sails on 1000 km 
scales \citep{Forward1984, Forward1985}. \citet{Landis1999} and \citet{Frisbee2009} found ways to reduce it dramatically 
to 1-10 km.  Systems were further optimized, with higher peak power of $\sim 10\,$TW and smaller vehicle size of $\sim 0.1-1\,$km for the sail, requiring $\sim 100\,$km antenna array aperture \citep{Dickinson2001, Long2011}.    Later concepts developed cost-optimized 
systems \citep{Benford2013}.

\subsection{Space solar power stations}

Space solar power stations, using microwave beams to efficiently transport 
power from solar cells in space to a planet's surface, are not likely to be observable. 
(Concepts for SSP vary from the microwave bands to lasers in the optical.) The 
beam must be carefully controlled to deliver power to the receiving rectifying 
antennas on the ground \citep{Mankins2014}.  Any side emissions are economic losses, 
therefore substantial measures would be taken to reduce side lobes to a minimum. 
 Further, the first several sidelobes are absorbed in the ground. The remaining 
side lobes are dispersed in angle so that the power density in the far field will 
be very low. For the worked example in Mankins \citep{Mankins2014, Dickinson2016}, 
the back lobe is down 40 dB relative to the $\sim 1\,$GW main beam. This 
is in contrast to power beaming transportation applications, in which the varying 
solid angle of the receiving spacecraft results in the main beam increasingly leaking 
around the edges of the vehicle being accelerated.

\section{PARAMETER SPACE OF POWER BEAMING OBSERVABLES}

We have surveyed 20 concepts, referenced in the above remarks, for power beaming 
systems which have sufficient detail to determine whether they could be observed. 
Tables~\ref{table1} and \ref{table2} summarize parameters of the power beaming applications discussed 
here, the power radiated, duration, and likely time for the radiation to repeat. 
The beamwidth is given by diffraction, $\Delta$$\theta$=2.44$\lambda$/\textit{D}, 
where $\lambda$ is the wavelength at the operating frequency and \textit{D} is the 
effective diameter of the radiating aperture, likely a phased array of either antennas 
or optics. 

\begin{figure*}
\epsscale{.80}
\plotone{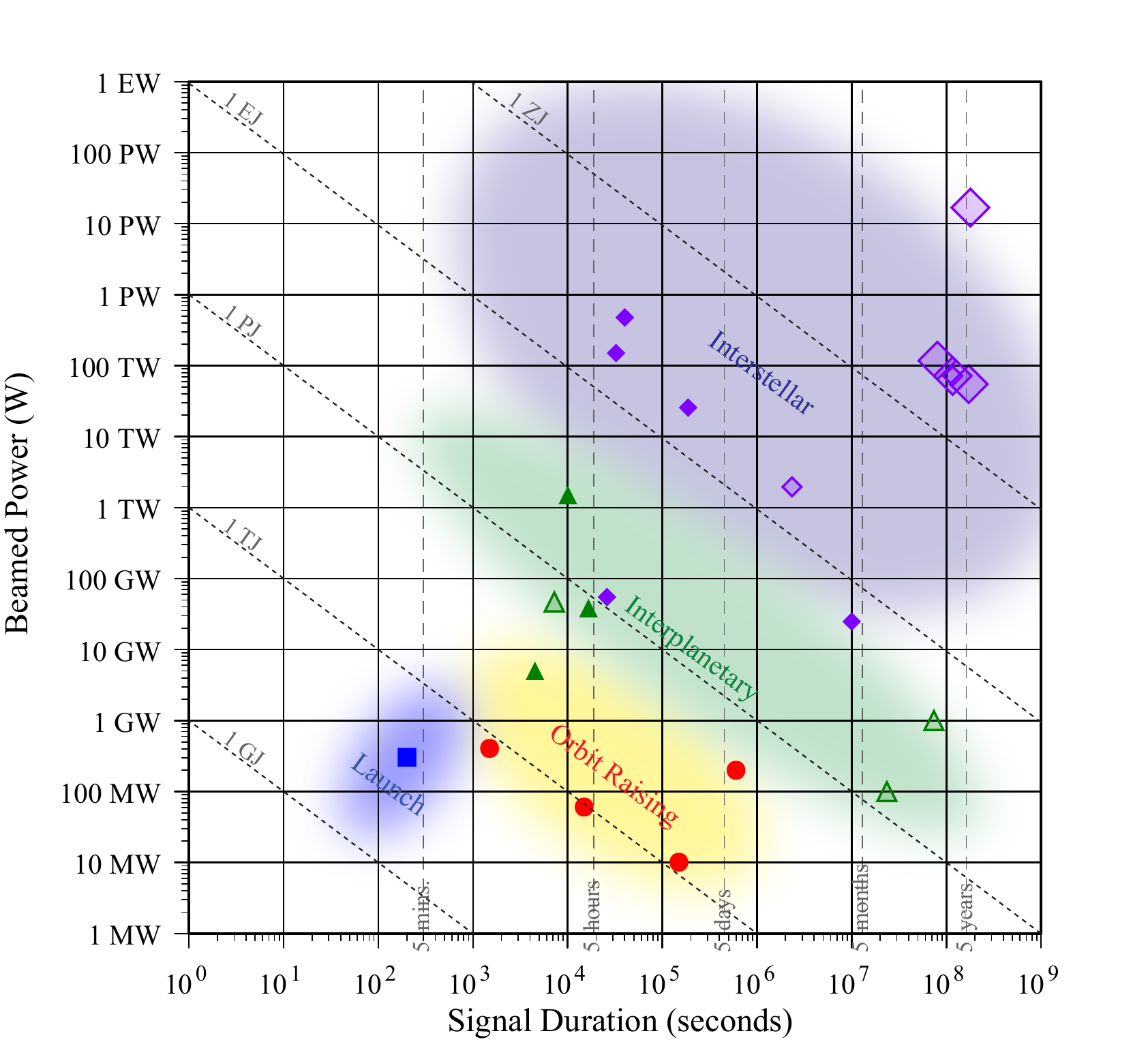}
\caption{Domains of beam power and duration of power beaming applications. Symbols indicate launch to orbit ($\blacksquare$), orbit raising (\mbox{\Large$\bullet$}), interplanetary ($\blacktriangle$), interstellar $0.1c$  ($\diamond$), and interstellar $0.5c$ (\mbox{\LARGE$\diamond$}). Data for the concepts are from the references. Solid symbols are for microwave and millimeter beaming, hollow symbols indicate visible/near-IR laser beams.\label{fig1}}
\end{figure*}

Table~\ref{table1} gives rough average powers and durations averaged over a given application. 
The power required by the applications varies by many orders of magnitude, with 
the launch to orbit and orbit raising application at levels below $\sim 1$ 
GW and the interplanetary and interstellar applications at far higher powers, into 
the TW and PW range. The increasingly energetic missions all require higher power 
and longer durations.  This corresponds to the velocities needed varying by four 
orders of magnitude. The repeat times also increase steadily as the energy requirement 
rises.  Figure~\ref{fig1} shows the power-duration parameter space. 

Aperture gain is set by the angular width of the beam $\Delta\theta$,
$G=\frac{4\pi}{(\Delta\theta)^2}$. The power 
density $S$ at range $R$ is determined by $W$, the effective isotropic radiated power 
(EIRP), which is the product of radiated peak power $P$ and aperture gain $G$,
\begin{equation}
W=PG \qquad\text{and}\qquad S=\frac{W}{4\pi R^2}.
\end{equation}

Spectral flux density, typically denoted in Janskys, is the power density divided 
by the bandwidth. While this is commonly used as the observed quantity 
in radio astronomy, we cannot know the bandwidth of an ETI transmitter. Consequently, 
in thinking about ETI power beaming emission we must deal with EIRP, not spectral 
flux density. Beaming power does not require or even necessarily benefit from 
narrow bandwidth; energy transference is what matters. To highlight this point, 
we have drawn diagonal lines of constant energy; there is a trend for applications 
of a certain type to follow these lines. For scale, the kinetic energy of a 5 ton 
vehicle moving at interplanetary speeds ($\sim 20$ km/s) is 1 TJ.

Beam widths and beam slew rates, the rate at which the beam moves to follow the 
spacecraft, and therefore sweeps past the observer, decline with power. The \textit{observation 
time} is the duration when the beam leakage could be detected. It is the beamwidth 
divided by the slew rate, $T=\Delta\theta/(d\phi/dt)$, 
is short and increases with higher power applications.

Table~\ref{table2} shows observables of power beaming at long range: slew rate, the EIRP and 
the observation time. Observers must be able to record transients over periods 
of order at least days.

The beam slew rate, $d\phi/dt$, is given by mission requirements. Slew 
rates are slow relative to planetary and stellar rotation rates. Observation times 
tend to be short, ranging from a few seconds to about 10 ms, a span of a couple 
of orders of magnitude.  The reason for that broad span is that sailship concepts 
proposed have velocities that vary by similar amounts. With a launch driven by 
an intense beam to arrive years later at a neighboring stellar system, the starship 
would be launched toward where the stellar system will be when the starship arrives. 
The ratio of the distance the star would move to the beam spot size is given by 
$v_s/(v_{ss} \Delta\theta)$, where $v_s$ 
is the average velocity of the star relative to stars on our stellar neighborhood, 
typically 20 km/sec, and $v_{ss}$ is the starship velocity.  For the 
starship concepts proposed, that ratio varies from $10^4$ to $10^7$. 
 The angle of the radiated beam with respect to the light path between the two stars is larger than the width of the beam. Thus, the beam is generally not observable from the target planetary system.

Very high-power devices might be located in space, so that atmospheric windows 
would not matter and frequency would depend upon the availability of efficient 
microwave, millimeter-wave or laser sources.  At present on Earth the most developed 
sources with high efficiencies and fairly low cost are in the microwave and millimeter-wave 
regime. ETI may well have far more advanced technology and be able to generate 
high power beams at any frequency.

\begin{deluxetable}{cccc}
\tablecolumns{4}
%\tabletypesize{\scriptsize}
\tablecaption{Representative Observable Parameters for \\ Applications of Power Beaming\label{table2}}
\tablewidth{0pc}
\tablehead{
\colhead{Application} & \colhead{Slew Rate} & \colhead{EIRP} & \colhead{Time} \\
\colhead{} & \colhead{$d\phi/dt$} & \colhead{$W=4\pi P/(\Delta\theta)^2$} & \colhead{$\Delta\theta/(d\phi/dt)$} \\
\colhead{} & \colhead{[rad/sec]} & \colhead{[W]} & \colhead{[sec]} }
\startdata
Launch to Orbit & $5\times 10^{-3}$ & $10^{19}$ & 0.04\\
Orbit Raising & $10^{-4}$ - $10^{-5}$ & $10^{16}$ & 3\\
Interplanetary & $7\times10^{-8}$ & $10^{25}$ & $0.04-0.4$\\
0.1 c Starship & 0 & $10^{32}$ & long\\
0.5 c Starship & 0 & $10^{38}$ & long
\enddata
\end{deluxetable}

\section{POWER BEAMING FROM KIC 8462852?}

Based on the above quantities, the recent report from the SETI Institute of radio 
observations of the anomalous star \objectname{KIC 8462852} has immediate implications \citep{Harp2016}.  That report concluded that, using the Allen Telescope Array, in 
the $1-10$ GHz microwave range, 1) no ``narrowband'' signals (1 Hz channels) were 
found above an EIRP of $4-7\times 10^{15}\,$W, and 2) no ``moderate band'' 
signals (100~kHz channels) were seen in  above an EIRP of $10^{19}$ W. 
 The observations spanned 2 weeks, observing half the time.

Comparing the reported thresholds set by the ATA observations to the power beaming 
applications summarized in Tables 1 and 2, the non-detection of leakage signals 
at their stated thresholds implies the following:

\begin{itemize}
\item The 1 Hz channels could see all the applications, but they are not seen.

\item Launch from a planetary surface into orbits is marginally detectable, at the threshold of the Allen Array for the 100\,kHz observations, if at the frequencies observed. Orbit raising, which requires lower power, is not detectable. 

\item Interplanetary transfers by beam-driven sails should be detectable in their 
observations, but are not seen. This is for both the 1 Hz and for the 100 kHz observations.

\item  Starships launched by power beams with beamwidths that we happen to fall within 
(to other solar systems, not our own) would be detectable, but are not seen.
\end{itemize}

In addition to radio measurements, an optical SETI measurement has been conducted towards \objectname{KIC 8462852} using the Boquete observatory \citep{Shuetz}. Its photomultiplier detector has a detection threshold of 67~photons/m$^2$ using a 25~ns gate time. Assuming that the signal to noise improves as the square root of bandwidth, for the times shown in the first three rows of Table~\ref{table2}, the detection limit in EIRP would be around $5\times 10^{26}\,$W. The longer times for starships would be implausible to detect with this technique, which does not support measurements of photon fluxes that are constant for more than a few seconds. Hence, none of the power beaming applications would be likely to have been detected if using visible photons.

These results must be qualified by noting:

\begin{itemize}
\item Power beaming is not an isotropic endeavor, and so the geometry of the transmitter 
and the intended recipient will produce a conjunction from our point of view only 
episodically. The observations were conducted for only a limited time and further 
observations would provide more stringent constraints. In general, the beamwidths in Table~\ref{table1} give an idea of the likelihood of intercepting power beaming leakage radiation.

\item The examples presented in Figure~\ref{fig1} that guide the estimates for observable parameters represent a broad summary of available studies; they are not, however, a coherent set of conceptual designs. They sample a range of assumptions and purposes, and were not necessarily optimized in any similar fashion. The shaded regions in the figure would improve from a more comprehensive exploration of the design space of beam-driven craft.

\item The powers and timescales presume launches for purposes similar to those studied, and for our planet and solar system. Substantially different applications, crafts, or payloads may yield different results, as would launches in a substantially different gravitational environment (for instance, if used commonly for transport around an asteroidal zone). It is possible to use a high powered laser to move solar system objects, such as diverting asteroids for planetary protection \citep{Lubin14} or evaporation and desorption of an comet's icy surface could produce a thrust to guide it into a planet (such as Mars, to produce lakes). 

\item For the radio measurements, even the ``moderate band'' observation is actually quite narrow compared with 
the kinds of sources that would be used in power beams, based on our current understanding 
of microwave physics. For the applications discussed here, the 100 kHz bandwidth 
observed would be about 10 to 100 millionths of the center frequency of the transmitter. 
 High-power devices using presently-understood physics are not designed for such 
narrow bandwidths. In microwave and millimeter wave devices on Earth, bandwidth is seldom a key parameter; other factors such as power are more important. Some examples: high-power gyrotrons are very narrowband devices.  They are highly overmoded, so have to be narrowband to avoid competition between modes, which reduces power. One $2\,$MW gyrotron operating at $140\,$GHz has a bandwidth of $70\,$MHz, which is 0.05\% bandwidth \citep{Thumm15}.  But 100 kW-class pulsed gyro-backward wave oscillators have up to 17\% bandwidth. Klystrons have bandwidth fractions of $\sim 0.1\%$. Consequently, future SETI observations should take such bandwidths into account.

\item The optimal radio frequencies we would presently use for power beaming are in the 
millimeter band, so are outside the microwave range the Allen Telescope Array observed. 
Similarly, power beaming using near-IR frequencies would be undetectable 
by the Boquete observatory.
\end{itemize}

Therefore, the \citet{Harp2016} and \citet{Shuetz} limited observation times and wavelength coverage are 
not sufficient to produce firm conclusions on power beaming from KIC 8462852. Most 
applications would be seen in the radio -- if transmissions were oriented in our direction at 
the proper time and at the frequencies observed -- but are not. More extensive observations should be made in more systematic studies of power beaming leakage, including observing at higher radio frequencies and for longer times at visible wavelengths.

\section{CONCLUSIONS AND IMPLICATIONS}

We have listed several classes of power beaming applications/missions, quantifying 
the principal observable parameters. Applying this reasoning to the recent observations 
of KIC~8462852, we conclude that if power beaming were in use at the time observed, 
generally pointed in our direction, and at frequencies between 1 and 10 GHz, that 
most power beaming applications would have been detectable. The nondetection provides 
a weak (owing to the caveat of the probabilities listed) rejection of the popular 
hypothesis that the system is inhabited by an advanced spacefaring extraterrestrial 
civilization.

As discussed above, the beaming power levels are high and transient and easily dwarf any ETI civilization's diffuse leakage to space \citep{Sullivan1978}. Power beaming described here is larger than that necessary for beaming systems for communication: $EIRP=10^{18}\,$W for a $1,000\,$ly-range beacon \citep{Benford3}. 

SETI programs could explore a different part of parameter space by observations suitable to finding leakage from power beams. Such beams would be visible over large interstellar distances. This implies a new approach to the SETI search: Instead of focusing on narrowband beacon transmissions generated by another civilization, look for more powerful beams with much wider bandwidth. This requires a new approach for their discovery by telescopes on Earth. Past SETI observations have been in the $1-10\,$GHz microwave band. For our atmosphere, future observations should look in bands where with lower oxygen and water vapor allow transmission: windows at $35\,$GHz, $70-115\,$GHz, $130-170\,$GHz and $200-320\,$GHz. And, of course, such transient sources require longer observing times.  A promising avenue is to revisit past observations of transient events, of which there are many, to look for patterns and identify as possible regions of the sky to emphasize.

Extraterrestrial intelligences would know their power beams could be observed. They could put a message on the power beam and broadcast it for our receipt at little additional energy or cost. By observing leakage from power beams we may well find a message embedded on the beam. That message may use optimized power-efficient designs such as spread spectrum and energy minimization \citep{Messerschmitt2012, Messerschmitt2015}.
 
If we build large power beaming systems in the future, we should be mindful of the possibilities of increased detectable leakage from Earth due to them. Such radiation may be a message, whether intentional or not.

%We have listed several classes of power beaming applications/missions, quantifying 
%the principal observable parameters. Applying this reasoning to the recent observations 
%of KIC~8462852, we conclude that if power beaming were in use at the time observed, 
%generally pointed in our direction, and at frequencies between 1 and 10 GHz, that 
%most power beaming applications would have been detectable. The nondetection provides 
%a weak (owing to the caveat of the probabilities listed) rejection of the popular 
%hypothesis that the system is inhabited by an advanced spacefaring extraterrestrial 
%civilization.
%
%As discussed above, the beaming power levels are high and transient and could easily 
%dwarf any ETI civilization's diffuse leakage to space. We should be mindful of 
%the possibilities of increased leakage from Earth in the future, if we build large 
%power beaming systems. 
%
%Extraterrestrial intelligences would know their power beams could be observed. 
%They could put a message on the power beam and broadcast it for our receipt at 
%little additional energy or cost. By observing leakage from power beams we may 
%well find a message embedded on the beam. That message may use optimized power-efficient 
%designs such as spread spectrum and energy minimization \citep{Messerschmitt2012, Messerschmitt2015}. If we in future build large power beaming systems we may well put messages on them. 

\acknowledgments

We are grateful for technical discussions with Kevin Parkin, Ian Morrison, David Messerschmitt, Gregory Benford, Manfred Thumm, and Gregory Nusinovich.


%% in parentheses, after the keyword, but they will not be verified.

{\it Facilities:} \facility{Allen Telescope Array}.



\clearpage

%% the file, uncomment the \plotone, \plottwo, and
%% If you need a layout that cannot be achieved with \plotone or
%% \plottwo, you can invoke the graphicx package directly with the
%% \includegraphics command or use \plotfiddle. For more information,
%% please see the tutorial on "Using Electronic Art with AASTeX" in the
%% documentation section at the AASTeX Web site, http://aastex.aas.org/
%%
%% The examples below also include sample markup for submission of
%% supplemental electronic materials. As always, be sure to check
%% the instructions to authors for the journal you are submitting to
%% for specific submissions guidelines as they vary from

%% This example uses \plotone to include an EPS file scaled to
%% 80% of its natural size with \epsscale. Its caption
%% has been written to indicate that additional figure parts will be
%% available in the electronic journal.

\clearpage

%% Here we use \plottwo to present two versions of the same figure,
%% one in black and white for print the other in RGB color
%% for online presentation. Note that the caption indicates
%% that a color version of the figure will be available online.
%%

%% If you are not including electonic art with your submission, you may
%% mark up your captions using the \figcaption command. See the
%% No more than seven \figcaption commands are allowed per page,
%% so if you have more than seven captions, insert a \clearpage

%% Tables should be submitted one per page, so put a \clearpage before

%% Two options are available to the author for producing tables:  the
%% deluxetable environment provided by the AASTeX package or the LaTeX
%% table environment.  Use of deluxetable is preferred.
%%

%% Three table samples follow, two marked up in the deluxetable environment,
%% one marked up as a LaTeX table.

%% In this first example, note that the \tabletypesize{}
%% command has been used to reduce the font size of the table.
%% We also use the \rotate command to rotate the table to
%% landscape orientation since it is very wide even at the
%% Note also that the \label command needs to be placed

%% This table also includes a table comment indicating that the full
%% version will be available in machine-readable format in the electronic

\clearpage


\begin{thebibliography}{}

\bibitem[Benford \& Benford(2006)]{Benford22006}Benford, G. \& Benford, J, 2006, Journal of the British Interplanetary Society, 59, 104

\bibitem[Benford \& Nissenson(2006)]{Benford2006}Benford, G. \& Nissenson, P., 2006, Journal of the British Interplanetary Society, 
59, 108

\bibitem[Benford(2008)]{Benford2008}Benford, J., 2008, IEEE Trans. Plasma Sci., 36, 569

\bibitem[Benford(2013)]{Benford2013}Benford, J. 2013, Journal of the British Interplanetary Society, 66, 85 

\bibitem[Benford, Benford \& Benford(2010)]{Benford3}Benford, J., Benford, G. \& Benford D., 2010, Astrobiology, 10, 475

\bibitem[Benford, Swegle \& Schamiloglu(2016)]{Benford2016}Benford, J., Swegle, J., Schamiloglu, E., 2016, High Power Microwaves, 3\textsuperscript{rd} ed., Taylor \& Francis, Boca Raton, FL

\bibitem[Billingham \& Benford(2014)]{Billingham2014}Billingham, J. \& Benford, J., 2014, Journal of the British Interplanetary Society, 67, 17

\bibitem[Brown(1992)]{Brown1992}Brown, W.C, 1992, IEEE Trans. Microwave Th. \& Tech., 40, 123

\bibitem[Cassenti(1982)]{Cassenti82}Cassenti, B. 1982, J. British Interplanetary Society, 35. 116 

\bibitem[Dickinson(2001)]{Dickinson2001}Dickinson, R. M., 2001, STAIF-2001, AIP Conf. Proc. 551, Albuquerque, NM

\bibitem[Dickinson(2016)]{Dickinson2016}Dickinson, R. M., 2016, private communication

\bibitem[Dyson(1982)]{Dyson82}Dyson, F. J., 1982, in {\it Extraterrestrials: where are they?} eds. M.H. Hart \& B. Zuckerman, 41, Pergamon Press, Oxford

\bibitem[Forward(1984)]{Forward1984}Forward, R. L., 1984, J. Spacecraft and Rockets, 21, 187

\bibitem[Forward(1985)]{Forward1985}Forward, R. L., 1985, J. Spacecraft and Rockets, 22, 345

\bibitem[Frisbee(2009)]{Frisbee2009}Frisbee, R., 2009, in Frontiers of Propulsion Science, Vol., ed. M. G. Millis and E. W. Davis, Prog. Astronautics and Aeronautics, 227, AIAA Press, Reston VA

\bibitem[Guillochon \& Loeb(2015)]{Guillochon2015}Guillochon, J. \& Loeb, A., 2015 ApJ Lett\textit{.} 811 L20

\bibitem[Harp et al.(2016)]{Harp2016}Harp, G. R. et al., arXiv:1511.01606v\emph{1} [astro-ph.EP]

\bibitem[Landis(1999)]{Landis1999}Landis, G., 1999, Journal of the British Interplanetary Society, 52, 420

\bibitem[Long(2011)]{Long2011}Long, K. F., 2011, Deep Space Propulsion, Springer, New York

%\bibitem[Landis(2000)]{Landis2000}Landis, G., 2000, AIAA-2000-3337, Fort Lauderdale, FL

\bibitem[Lubin(2014)]{Lubin14}Lubin, P et al, 2014, Optical Engineering, 53, 025103

\bibitem[Mankins(2014)]{Mankins2014}Mankins, J., 2014, The Case for Space Solar Power, Virginia Edition, Raleigh, NC.

\bibitem[Matloff(2005)]{Matloff05}Matloff, G.L. 2005, {\it Deep Space Probes}, 2nd Edition, Springer-Verlag , NY

\bibitem[Messerschmitt(2012)]{Messerschmitt2012}Messerschmitt, D., 2012, Acta Astronautica, 81, 227

\bibitem[Messerschmitt(2015)]{Messerschmitt2015}Messerschmitt, D., 2015, Acta Astronautica, 107, 20

\bibitem[Meyer et al.(1985)]{Meyer1985}Meyer, T., McKay, C., McKenna, P., \& Pryor, W., 1985, The Case for Mars II, San Diego, CA, Univelt, Inc., 419

\bibitem[Moeckel(1972)]{Moeckel72}Moeckel, W.E., 1972, J. Spacecraft and Rockets, 9, 863

\bibitem[Parkin et al.(2003)]{Parkin2003}Parkin, K. L. G. et al., 2003, Proc. ISBEP2, Sendai, Japan, AIP, 324. 

\bibitem[Shuetz et al.(2016)]{Shuetz}Schuetz, M. et al. 2016, ApJL, submitted

\bibitem[Sullivan(1978)]{Sullivan1978}Sullivan W.T. III, et al., 1978, \textit{Science }199, 377 

\bibitem[Thumm(2015)]{Thumm15}Thumm, M., 2015, private communication

\end{thebibliography}
\end{document}